# GENERALIZED HEISENBERG UNCERTAINTY RELATION IN SPERICAL COORDINATES

ANZOR KHELASHVILI

*Inst.of High Energy Physics,,Iv.Javakhihvili Tbilisi State University*
*G.Danelia Str. 10 , Tbilisi 0109, Georgia*
anzor.khelashvili@tsu.ge

TEIMURAZ NADAREISHVILI

*Inst.of High Energy Physics,,Iv.Javakhihvili Tbilisi State University*
*G.Danelia Str. 10 , Tbilisi 0109, Georgia*
*Faculty of Exact and Natural Sciences, Iv.Javakhihvili Tbilisi State University,*
*Chavchavadze Ave 3, Tbilisi 0179,Georgia*
teimuraz.nadareishvili@tsu.ge

Following to the Weil method we generalize the Heisenberg-Robertson uncertainty relation for arbitrary two operators. Consideration is made in spherical coordinates, where the distant variable $r$ is restricted from one side, $0 \leq r < \infty$. By this reason accounting of suitable boundary condition at the origin for radial wave functions and operators is necessary. Therefore, there arise extra surface terms in comparison with traditional approaches. These extra terms are calculated for various solvable potentials and their influence is investigated. At last, the time-energy uncertainty relations are also analyzed. Some differences between our approach and that, in which a direct product for separate variances were considered, is discussed.

.

## 1. Introduction

From time to time there emerge papers, in which the well-known Heisenberg – Robertson (HR) uncertainty relations are modified by one or other reasons. Mainly it is associated with strong mathematical properties of considered operators.

The uncertainty relation forms the basis of entire quantum theory. The ordinary proof of the HR uncertainty relation for coordinate and corresponding momentum is based on the Cauchy-Schwarz inequality, which is a general property of Hilbert space. For a pair of operators $\hat{A}$ and $\hat{B}$, acting on the state $|\Psi\rangle$, the Schwartz inequality gives

$$\langle \hat{A}\Psi | \hat{A}\Psi \rangle \langle \hat{B}\Psi | \hat{B}\Psi \rangle \geq |\langle \hat{A}\Psi | \hat{B}\Psi \rangle|^2 \qquad (1)$$

Transition to the uncertainty relation for *variance (or mean square deviation)* $|\Delta A \Delta B|$ where $\Delta A \equiv \hat{A} - \langle \hat{A} \rangle$, requires to impose some fundamental limitations on operators under consideration, such as hermiticity, unique domains, self-adjointness etc. After that such limitations are applied, the general uncertainty relation for any such operators reduces to

$$|\langle \Delta A \Delta B \rangle|^2 \geq \frac{1}{4}|\langle [\hat{A},\hat{B}] \rangle|^2 \qquad (2)$$

which is a source of uncertainty inequalities for various operators.

However, it was pointed out earlier [1,2], that the commutation relation

$$[\hat{A},\hat{B}] = i\hbar \hat{C} \qquad (3)$$

for quantum observables $\hat{A}, \hat{B}$ and $\hat{C}$ by itself does not imply the uncertainty relation

$$\langle \Delta A \Delta B \rangle \geq \frac{\hbar}{2}|\langle \hat{C} \rangle| \qquad (4)$$

for all physically interesting states. There are examples showing that the uncertainty relation (2) for any two observables $\hat{A}$ and $\hat{B}$ is not valid in such a generality. For example, the uncertainty relation for $\hat{A}$ and $\hat{B}$ is usually written in the form (2), or more explicitly, as

$$\Delta_\psi A \Delta_\psi B \geq \frac{\hbar}{2}|(\psi, i[\hat{A},\hat{B}]\psi)| \qquad (5)$$

where $(\Delta_\psi A)^2 = \|(\hat{A} - \langle \hat{A}_\psi \rangle)\psi\|^2$, with $\langle \hat{A}_\psi \rangle = (\psi, \hat{A}\psi)$ and likewise for $\hat{B}$. This definition is used in many papers [3-5]. Thus, the left – hand side of relation (5) is defined for $\psi \in D(A) \cap D(B)$. On the other hand, the right –side is only defined on the subspace $D([A,B]) = D(AB) \cap D(BA)$, which is much smaller in general.

However, $\hat{A}$ and $\hat{B}$ being self-adjoint, relation (5) can be rewritten in the form [2,3]

$$\Delta_\psi A \Delta_\psi B \geq \frac{\hbar}{2} |i(A\psi, B\psi) - i(B\psi, A\psi)| \qquad (6)$$

Where the domain of the right-hand side now coincides with that of the left-hand side, $D(A) \cap D(B)$, i.e., the product of uncertainties for two observables $\hat{A}$ and $\hat{B}$ is not determined by their commutator, but by hermitian sesquilinear form [2,3]

$$\Phi_{A,B}(f,g) = i(Af, Bg) - i(Bf, Ag) \qquad (7)$$

for all $f, g \in D(A) \cap D(B)$.

As it is mentioned in [3], this modification is not simply a cosmetic one. In cases, when $\hat{A}$ or $\hat{B}$ are operators of differentiation, the surface terms occurring upon integration by parts do not vanish in general and contribute to the uncertainty relations.

A well-known manifestation of this specific situation occurs in 3-dimensions, when spherical coordinates are used. One of the basic coordinates, namely, a distance, is restricted by a half-line. Therefore, we have to be careful in calculation of surface integrals, which, because of boundary conditions at the origin, may also give the nontrivial contributions. Moreover, the singularity of considered operators can also contribute to uncertainty inequalities

While the strong mathematical considerations are most powerful [3], the explicit calculations have more transparency and simplicity. It was already demonstrated clearly in number of papers [6,7], concerning to the Ehrenfest theorem, published recently. We are inclined to think that a similar simplification may be also occurred in case of uncertainty relations.

The aim of this paper is a consideration of a Heisenberg's uncertainty relations in 3-dimensional quantum mechanics (Schrodinger theory), using spherical coordinates. The point is that in this system the boundary behavior of wave functions and singularity of operators play important role – there appear extra terms owing to this behavior, which contribute non-trivially to the surface integrals. These terms may be calculated in explicit form.

The paper is organized in the following way: First of all, we give a general consideration of the Weil method for deriving uncertainty relation for arbitrary two operators $\hat{A}$ and $\hat{B}$, taking into account that, owing to boundary behavior of radial wave functions some additional surface terms may arise. After that by using the well-known technique, corresponding uncertainty inequality is derived for the central symmetric problem. Then we derive the general inequality between $r^2$ and $p^2$ in case of regular boundary condition (regular potentials), thereafter we calculate the right-hand side of uncertainty relation for various solvable potentials in explicit forms, in particular, for any operator with the Hamiltonian. Finally, some brief comments are given for the time-energy uncertainty relations in the light of the recent publications.

## 2. General consideration of the Heisenberg-Robertson (HR) relation

In fact, there are known various ways for deriving HR relation. Though they differ from each other's, but all of them must be consistent with the fundamental Heisenberg inequality. Generalizations of HR relation contain some additional expressions, which can appear in course of partial integration as surface terms.

First of all, we chose a general derivation of the uncertainty relation following to a Weil method, which can be found in number of textbooks. Below we follow to Davidov's book [8] (see, also the exercise book of Galitski [9])

Let us suppose that the commutation relation between two Hermitian operators $\hat{A}$ and $\hat{B}$ has the form (3), where $\hat{C}$ is a hermitian operator too (with certain restrictions on the wave functions). Consider the integral

$$J(\alpha) = \int \left| \left( \alpha \hat{A}_1 - i\hat{B}_1 \right) \psi \right|^2 d\tau \geq 0 \qquad (8)$$

where $\hat{A}_1 = \hat{A} - a$, $\hat{B}_1 = \hat{B} - b$ with $\alpha, a, b$ are some real parameters. We rewrite the integral in the form

$$J = \int \left( \left( \alpha \hat{A}_1 - i\hat{B}_1 \right) \psi \right)^* \left( \alpha \hat{A}_1 - i\hat{B}_1 \right) \psi d\tau = \int \int \left( \alpha \hat{A}_1 + i\hat{B}_1 \right) \psi^* \left( \alpha \hat{A}_1 - i\hat{B}_1 \right) \psi d\tau \qquad (9)$$

Despite the fact that $\hat{A}$ and $\hat{B}$ are Hermitian operators, we write down the Hermiticity condition with some care, introducing the following notations

$$\int (\hat{A}_1 \psi)^* \hat{A}_1 \psi d\tau = \int \psi^* \hat{A}_1^2 \psi d\tau + Q_1 \qquad (10)$$

$$\int (\hat{B}_1 \psi)^* \hat{B}_1 \psi d\tau = \int \psi^* \hat{B}_1^2 \psi d\tau + Q_2 \qquad (11)$$

$$\int (\hat{A}_1 \psi)^* \hat{B}_1 \psi d\tau = \int \psi^* \hat{A}_1 \hat{B}_1 \psi d\tau + X \qquad (12)$$

$$\int (\hat{B}_1 \psi)^* \hat{A}_1 \psi d\tau = \int \psi^* \hat{B}_1 \hat{A}_1 \psi d\tau + Y \qquad (13)$$

Here $Q_1, Q_2, X, Y$ are surface terms, remaining after integration by parts, which may be non-vanishing on the boundaries. They show deviations from (6). Exactly these terms can modify the uncertainty relations.

Let us drown an analogy between above decomposition and that given in paper [5]. They introduced the *expectation commutator* in the following way:

$$\hat{A} \# \hat{B} := \left\langle \hat{A}\psi \middle| \hat{B}\psi \right\rangle - \left\langle \psi \middle| \hat{A}\hat{B}\psi \right\rangle \qquad (14)$$

and instead of (5), considered new uncertainty relation

$$\Delta A \Delta B \geq \frac{1}{2} \left| \hat{A} \# \hat{B} - \hat{B} \# \hat{A} + \left\langle [\hat{A}, \hat{B}] \right\rangle \right| \qquad (15)$$

It is easily to realize that the extra term here corresponds to ours $X - Y$. Therefore, our consideration is basically equivalent to that of [5], but a careful and straightforward calculation often has an advantage. For example, new terms $Q_1$ and $Q_2$ may also appear.

Taking into account notations above, the sought-for integral (8) can be rewritten in the following form (assumed the normalization of the state $\psi$)

$$J = \int \psi^* \left( \alpha^2 \hat{A}_1^2 - i\alpha [\hat{A}_1, \hat{B}_1] + \hat{B}_1^2 + Q_1 + Q_2 + i\alpha(Y - X) \right) \psi d\tau = \\ = \alpha^2 \langle \hat{A}_1^2 \rangle - i\alpha \langle [\hat{A}_1, \hat{B}_1] \rangle + \langle \hat{B}_1^2 \rangle + Q_1 + Q_2 + i\alpha(Y - X) \tag{16}$$

and after taking into account the commutation relations $[\hat{A}_1, \hat{B}_1] = [\hat{A}, \hat{B}] = i\hbar \hat{C}$, we find

$$\alpha^2 \langle \hat{A}_1^2 \rangle + \alpha \langle \hbar \hat{C} + i(Y - X) \rangle + \langle \hat{B}_1^2 \rangle + Q_1 + Q_2 \geq 0 \tag{17}$$

Now from positive definiteness of square trinomial, we conclude that

$$\langle (\hat{A}_1^2) \rangle \left[ \langle \hat{B}_1^2 \rangle + Q_1 + Q_2 \right] \geq \frac{1}{4} \langle \hbar \hat{C} + i(Y - X) \rangle^2 \tag{18}$$

If we now return to above introduced notations and suppose in addition that $a = \langle A \rangle$ and $b = \langle B \rangle$, it follows generalized Heisenberg relation

$$\langle (\hat{A} - \langle \hat{A} \rangle)^2 \rangle \left[ \langle (\hat{B} - \langle \hat{B} \rangle)^2 \rangle + Q_1 + Q_2 \right] \geq \frac{1}{4} \langle \hbar \hat{C} + i(Y - X) \rangle^2 \tag{19}$$

or

$$\langle (\Delta A)^2 \rangle \left[ (\Delta B)^2 + Q_1 + Q_2 \right] \geq \frac{1}{4} \langle \hat{C} + i(Y - X) \rangle^2 \tag{20}$$

where

$$\langle (\Delta A)^2 \rangle = \langle (A - \langle A \rangle)^2 \rangle = \langle \hat{A}^2 \rangle - \langle \hat{A} \rangle^2 \tag{21}$$

Using the traditional expression

$$\delta A = \sqrt{\langle (\Delta A)^2 \rangle} \tag{22}$$

for a mean square-root deviation or variance [26], and likewise for $B$, the uncertainty inequality (20) takes the form

$$(\delta A)\sqrt{(\delta B)^2 + Q_1 + Q_2} \geq \frac{1}{2} \left| \langle \hat{C} + i(Y - X) \rangle \right| \tag{23}$$

Some comments are now in order here.

1. This relation reduces to the usual one (2), when $Q_1 = Q_2 = X = Y = 0$, i.e., when all considered operators are *self-adjoint*. For physical operators more strict conditions are needed: it is necessary that not only wave function $\psi$, but also $\hat{B}\psi$ belongs to the appropriate domain, in which $\hat{A}$ is hermitian (and similarly, $\hat{A}\psi$ must remain in the domain of wavefunctions, where $B$ is hermitian). However, in addition, if the area is limited and boundary conditions are also imposed, it may be that above mentioned restrictions fail. Only detailed calculation can shed light. In general, uncertainty product $\delta A \delta B$ does not always separate as a factor. As a rule, it appears in combination with $Q_1$ or $Q_2$. In this respect the Weil method, as such, is inconvenient to use, at a first glance. But the same happen in using other known methods, which are described in various textbooks [10,11], if the product of operators should be represented as above, Eqs. (10)-(13).

   As we will see below, explicit calculation gives that the additional terms from Eq. (21) *do not always disappear.* Only in cases, when additional terms are absent, the ordinary HR uncertainty relation (4) follows. The inequality (4) has been executed in arbitrary state for two observables, operators of which do not necessary commute.

2. When operators commute, $C = 0$ and, therefore, the two physical quantities are measurable simultaneously, i.e., one obviously can take $\delta A = \delta B = 0$. Then it follows from (20) the true inequality, $0 \geq -\frac{1}{4}\langle (Y-X)\rangle^2$. Hence (23) also contains a case of simultaneous measurement of two physical quantities.

3. It may happen that not all surface terms are zero simultaneously. As we will see below it depends on the singular character of considered operators at the boundaries.

4. If $Q_1 + Q_2 < 0$, then $(\delta B)^2$ is constrained from below, $(\delta B)^2 > |Q_1 + Q_2|$. In contrast, when $Q_1 + Q_2 > 0$, then $(\delta A)^2$ should be constraint

$$\delta A \geq \frac{1}{2}|\langle \hbar C + i(Y-X)\rangle| \frac{1}{\sqrt{Q_1 + Q_2}} \qquad (24)$$

So, one can establish which physical quantity will be constraint from below.
 In the following we get examples of application of (20) and/or (23).

As usual, minimization of uncertainty relation corresponds to the sign of equality in (20) and (23). Indeed, it we take $\hat{A} = x, \hat{B} = p_x$, then in this case $Q_1 = Q_2 = X = Y = 0$ and choose $a = x_0, b = p_0$ the explicit form of $\psi$ saturated (23) is as follows

$$\psi(x) = \left[2\pi\langle(\Delta x)^2\rangle\right]^{-1/4} \exp\left\{-\frac{(x-x_0)^2}{4\langle(\Delta x)^2\rangle} + \frac{ip_0 x}{\hbar}\right\} \quad (25)$$

In this state the uncertainties product is minimal

$$(\delta x)^2(\delta p)^2 = \langle(\Delta x)^2\rangle\langle(\Delta p_x)^2\rangle = \frac{\hbar^2}{4} \quad (26)$$

**3. More applications - Uncertainty relations for central symmetric problems**

The most papers are devoted to the angular momentum-angle uncertainty relation [1, 2, and 4]. We considered this example in [12] as well. Applying above formulas, we have improved results, known in current literature [1,2] and [4] and derived the consisting result. Below we consider only some problems, concerning to the cases of central symmetric Schrodinger equation. Similar problems are considered in number of papers [13, 14] in 3-dimensions, as well as in more dimensions [15-18]. But in the most of them only simple uncertainty products are studied for cases of solvable potentials in the Schrodinger equation – mainly for Coulomb, Harmonic oscillator or spherical wall – and the minimum of this product is established. As we have seen above, product of uncertainties contains the average values of appropriate commutators. Nobody considered these averages. But by our opinion, it is interesting to compare estimated values to the ordinary uncertainty products.

In [6, 7] we have described Ehrenfest and Hypervirial theorems for central symmetric potentials. Now let us consider the cases when operators $\hat{A}$ and $\hat{B}$ depend on radial distance and correspond to time-independent physical observables. The Hamiltonian is

$$\hat{H}_R = \frac{\hbar^2}{2m}\left(-\frac{d^2}{dr^2} - \frac{2}{r}\frac{d}{dr}\right) + \frac{\hbar^2 l(l+1)}{2mr^2} + V(r) \quad (27)$$

,
where the potential $V(r)$ is either regular

$$\lim_{r\to 0} r^2 V = 0 \quad (28)$$

or soft-singular

$$\lim_{r\to 0} r^2 V \to \pm V_0 = const; V_0 > 0 \quad (29)$$

With the following regular boundary behaviour for total radial function [6] $\psi = R(r)Y_{lm}(\theta,\varphi)$, correspondingly

$$\lim_{r \to 0} R = Cr^l \tag{30}$$

for regular potential, or for soft-singular potential

$$\lim_{r \to 0} R = a_{st} r^{-\frac{1}{2}+P} \tag{31}$$

where

$$P = \sqrt{\left(l+\frac{1}{2}\right)^2 - \frac{2mV_0}{\hbar^2}} \tag{32}$$

Our anxiety is the relation between pair $\hat{A} = r^2$ and $\hat{B} = p^2$. Both of them are Hermitian operators. For calculation of their commutator, we make use an obvious relation

$$\hat{p}^2 = \hat{p}_r^2 + \frac{\hat{l}^2}{r^2} \quad , \tag{33}$$

where $\hat{l}^2$ is an orbital momentum operator and $\hat{p}_r$ is a Hermitian radial component of linear momentum

$$\hat{p}_r = -i\hbar \frac{1}{r}\frac{d}{dr} = -i\hbar\left(\frac{d}{dr} + \frac{1}{r}\right) \quad , \tag{34}$$

which obeys the following canonical commutation relation $[r, \hat{p}_r] = i\hbar$. Now taking into account that $r^2$ is a 3-dimensional scalar and commutes with the angular momentum operator, we deduce $[r^2, p^2] = [r^2, p_r^2]$.

The following steps are straightforward and one can obtain

$$[r^2, p^2] = 4\hbar^2\left(r\frac{d}{dr} + \frac{3}{2}\right) \tag{35}$$

Moreover, because $r^2$ is a multiplication operator, according to definitions (10), (12) we conclude that $Q_1 = X = 0$. As for $Q_2$, we use the relation (11) and obtain

$$\int_0^\infty (p^2 R)^* p^2 R r^2 dr = \int_0^\infty R^* p^4 R r^2 dr + Q_2 \tag{36}$$

where

$$Q_2 = 2m\hbar^2 a_{st}^2 \lim_{r \to 0} V' r^{2P+1} \tag{37}$$

In case of (29) soft singular potentials this term becomes

$$Q_2 = \mp 4 a_{st}^2 m\hbar^2 V_0 \lim_{r \to 0} r^{2P-2} \tag{38}$$

Therefore, it survives only for $P = 1$, when

$$Q_2 = \mp 4a_{st}^2 m\hbar^2 V_0 \tag{39}$$

## 3.1. Regular potentials

For (28) regular potentials [6] $P = l + 1/2$ and we must distinguish two possible cases, according to behavior of potential

$$V = -\frac{\alpha}{r^\beta} \tag{40}$$

(i) $0 < \beta < 1$

In this case, according to (37), we have

$$Q_2 = 2m\hbar^2 \beta\alpha a_{st}^2 \lim_{r\to 0} r^{2l+1-\beta}, \tag{41}$$

and we see, that $Q_2 = 0$

(i) On the other hand, when $0 \leq \beta < 2$, $Q_2$ survives only if $2l+1 = \beta$, in such case

$$Q_2 = 2m\hbar^2 \alpha a_{st}^2 (2l+1) \tag{42}$$

Otherwise $Q_2 = 0$, when $2l+1 > \beta$ and it diverges, if $2l+1 < \beta$. Nevertheless, for $l = 0$ it follows $Q_2 \neq 0$ (when $\beta = 1$) and we conclude that among regular potentials only for Coulomb one exists nonzero $Q_2$ given by (42). As regards of $a_{st}$ - it depends on the behavior of wave function at the origin for particular potentials. Also $Y = -2\hbar^2 a_{st}^2 \lim_{r\to 0} r^{2P+2} = 0$ as $P > 0$.

Collecting all above found relations and take $\hat{A} = r^2$ and $\hat{B} = p^2$, the general uncertainty relation takes the form

$$\left\langle (\Delta r^2)^2 \right\rangle \left( \left\langle (\Delta p^2)^2 \right\rangle + Q_2 \right) \geq 4\hbar^2 \left\langle \left( r\frac{d}{dr} + \frac{3}{2} \right) \right\rangle^2 \tag{43}$$

Or

$$\delta r^2 \sqrt{(\delta p^2)^2 + Q_2} \geq 2\hbar \left| \left\langle \left( r\frac{d}{dr} + \frac{3}{2} \right) \right\rangle \right| \tag{44}$$

where

$$\delta r^2 = \sqrt{\left\langle (\Delta r^2)^2 \right\rangle} = \sqrt{\left\langle (r^2 - \langle r^2 \rangle)^2 \right\rangle} = \sqrt{\langle r^4 \rangle - \langle r^2 \rangle^2} \tag{45}$$

$$\delta p^2 = \sqrt{\left\langle (\Delta p^2)^2 \right\rangle} = \sqrt{\left\langle (p^2 - \langle p^2 \rangle)^2 \right\rangle} = \sqrt{\langle p^4 \rangle - \langle p^2 \rangle^2} \tag{46}$$

This form of uncertainty relation differs from that, known in current literary. More familiar relation is [14, 19]

$$\Delta r \Delta p \geq \hbar \left( l + \frac{3}{2} \right) \tag{47}$$

However, as we'll see below, result depends on other characteristics of dynamics and is sensitive to other quantum numbers, i.e., has no universal character.

### 3.2. Calculation of r.-h.-s. of eq (44) for various solvable potentials

Let us calculate r.-h.-side of Eq. (44) for the attractive Coulomb potential, which has a form

$$V = -\frac{e^2}{r} \tag{48}$$

The wave function is [20]

$$R_{nl} = C_{nl} r^l e^{-\frac{Br}{2n}} F\left(-n_r, 2l+2; \frac{Br}{n}\right) \tag{49}$$

where

$$C_{nl} = \frac{B^{3/2}}{\sqrt{2} n^2 (2l+1)!} \sqrt{\frac{(n+l)!}{n_r!}} \left(\frac{B}{n}\right)^l \tag{50}$$

Is a normalization constant and

$$n = n_r + l + 1;\; B = \frac{2me^2}{\hbar^2} = \frac{2}{a_0} \tag{51}$$

where $a_0 = \frac{\hbar^2}{me^2}$ is Bohr's radius.

Our aim now is a calculation of average value $\left\langle r \frac{d}{dr} \right\rangle$. Derivative of hypergeometric function $F$, which occurs in the calculation, can be replaced by the known expression [21]

$$F'(a,b;x) = \frac{a}{b} F(a+1, b+1; x) \tag{52}$$

and then appearing one of the integral

$$I = -C_{nl}^2 \frac{B}{n} \frac{n_r}{(2l+2)} \int_0^\infty r^{2l+3} e^{-\frac{Br}{n}} F\left(-n_r, 2l+2; \frac{Br}{n}\right) F\left(-n_r+1, 2l+3; \frac{Br}{n}\right) dr \tag{53}$$

we transform according to the Suslov's relation for reducing of hypergeometric function in terms of the Laguerre polynomials [22]

$$F(-n, \beta+1; x) = \frac{n!\Gamma(\beta+1)}{\Gamma(\beta+1+n)} L_n^\beta(x) \qquad (54)$$

Then, it follows for (53) that

$$I = -C_{nl}^2 \frac{n_r}{2l+2} \frac{n_r!\Gamma(2l+2)}{\Gamma(2l+2+n_r)} \frac{(n_r-1)!\Gamma(2l+3)}{\Gamma(2l+3+n_r)} \left(\frac{n}{B}\right)^{2l+3} \int_0^\infty x^{2l+3} e^{-x} L_{n_r}^{2l+1}(x) L_{n_r-1}^{2l+2}(x) dx \qquad (55)$$

The last integral also is known from the formula for product of Laguerre polynomials [22]

$$J_{nms}^{ab} = \int_0^\infty e^{-x} x^{a+s} L_{n_1}^a(x) L_m^b(x) dx = (-1)^{n_1-m} \frac{\Gamma(a+s+1)\Gamma(b+m+1)\Gamma(s+1)}{m!(n_1-m)!\Gamma(b+1)\Gamma(s-n_1+m+1)} {}_3F_2\left(\begin{array}{c}-m, s+1, b-a-s\\b+1, n_1-m+1\end{array};1\right) \qquad (56)$$

where ${}_3F_2\left(\begin{array}{c}-m, s+1, b-a-s\\b+1, n_1-m+1\end{array};1\right)$ is a generalized Hypergeometric function at $x=1$.

A comparison of previous two forms gives
$$a = 2l+1; s = 2; m = n_r - 1 \qquad (57)$$

and therefore, we find

$$_3F_2\left(\begin{array}{c}-n_r+1, 3, -1\\2l+3, 2;\end{array};1\right) = 1 + \frac{3(1-n_r)}{2(2l+3)} \qquad (58)$$

Finally we have

$$I = C_{nl}^2 \left(\frac{n}{B}\right)^{2l+3} \frac{n_r 2n_r!\Gamma(2l+2)\Gamma(2l+4)}{\Gamma(n_r+2l+3)} \left[1 + \frac{3(1-n_r)}{2(2l+3)}\right] \qquad (59)$$

All remaining integrals can be performed in a perfect analogy, applying some table integrals from [22]. Operating to the same direction, we find the net result

$$\left\langle r\frac{d}{dr} + \frac{3}{2}\right\rangle = \frac{3}{2} + \frac{1}{2n}\left\{-3(l+1) + n_r\left[\frac{2(l+1)(4l+9-3n_r)}{2l+2+n_r} - 3n_r - 4l - 6\right]\right\}; \qquad (60)$$
$$n = n_r + l + 1$$

You can see evidently that for $n_r = 0$ and arbitrary $l$, this expression vanishes! K.Urbanowski took attention on such a fact [23]. He argued that for some state and for some non-commuting observables $\hat{A}$ and $\hat{B}$ may happen that $\langle \psi|[A,B]|\psi\rangle = 0$. He constructed two suitable examples too [24]. In our case we have analogous situation. Operators under consideration $r^2$ and $p^2$ are non-commuting, moreover their commutator is a nontrivial operator again. So, its average is zero in a node less state.
We derived above that for the Coulomb potential $Q_2 \neq 0$, specifically, $a_{st}^2 = C_{nl}^2$ is given by Eq. (50). Therefore,

$$a_{st}^2 = \frac{4}{(n_r+1)^3 a_0^3} \tag{61}$$

and

$$Q_2 = 2m\hbar^2 e^2 a_{st}^2 = \frac{8m\hbar^2 e^2}{(n_r+1)^3 a_0^3} > 0 \tag{62}$$

Hence the uncertainty relation takes the form (for $l = 0$)

$$\delta r^2 \sqrt{(\delta p^2)^2 + \frac{8m\hbar^2 e^2}{(n_r+1)^3 a_0^3}} \geq 2\hbar \left| \frac{3}{2} + \frac{1}{2n}\left\{-3 + n_r\left[\frac{18-6n_r}{2+n_r} - 3n_r - 6\right]\right\}\right|; \quad n = n_r+1, n_r = 0,1,2,... \tag{63}$$

While in case of nonzero orbital momentum ($l > 0$), because $Q_2 = 0$, we have

$$\delta r^2 \delta p^2 \geq 2\hbar \left|\frac{3}{2} + \frac{1}{2n}\left\{-3(l+1) + n_r\left[\frac{2(l+1)(4l+9-3n_r)}{2l+2+n_r} - 3n_r - 4l - 6\right]\right\}\right|; n = n_r+l+1 \tag{64}$$

The r.-h.-side of this relation vanishes for node less state $n_r = 0$.

It is worthwhile to note that in [25], the following relation was derived by production of $\langle r^2 \rangle$ and $\langle p^2 \rangle$ for the Coulomb potential

$$\langle r^2 \rangle \langle p^2 \rangle = \frac{\hbar^2}{2}\left[5n^2 - 3l(l+1) + 1\right]; \quad n = n_r + l + 1 \tag{65}$$

The same expression was derived in [15] for arbitrary spatial dimensions. We see, that this expression never vanishes. So, there appear a difference between explicit calculation and a direct multiplication approach.

Moreover, as well as $Q_2 > 0$ (see, Eq. (62)), one can consider $\Delta p \to 0$ in (59), i.e. to measure linear momentum precisely (exactly), then a spread of coordinate becomes bounded from below:

$$\delta r^2 \geq 2\hbar \left\langle \left( r\frac{d}{dr} + \frac{3}{2} \right) \right\rangle \frac{1}{\sqrt{Q_2}} \tag{66}$$

This case formally resembles the situation in the relativistic area [26], when the measurement of momentum is possible precisely, but not the coordinates. We briefly consider other potentials using above developed strategy in the next section.

### 3.3. Other potentials

Consider the following *soft-singular potential*:

$$V = -\frac{V_0}{r^2} + \alpha r^k; \quad V_0 > 0; \quad k > -2 \tag{67}$$

In this case Eq. (36) gives

$$Q_2 = 4m\hbar^2 V_0 a_{st}^2 \lim_{r \to 0} r^{2P-2} \tag{68}$$

and a non-vanishing $Q_2$ follows, when $P = 1$

$$Q_2 = 4m\hbar^2 a_{st}^2 V_0 \tag{69}$$

We see that $Q_2 > 0$ and the result is like of previous one.

- *The valence electron model potential* is

$$V = -\frac{\alpha}{r} - \frac{V_0}{r^2}; \quad V_0 > 0; \quad \alpha > 0 \tag{70}$$

Using solutions, derived in [6] and repeating all procedures, exploited above, one finds for this potential that

$$\left\langle r\frac{d}{dr} + \frac{3}{2} \right\rangle = \left| \frac{3}{2} + \frac{1}{(2n_r + 2P + 1)} \left\{ -\frac{3}{2}(2P+1) + n_r \left[ \frac{(2P+1)[2(2P+2) + 3(1-n_r)]}{1 + 2P + n_r} - 4P - 3n_r - 4 \right] \right\} \right|$$

(71)

As in previous case, here at $n_r = 0$ and for arbitrary $l$ it follows again vanishing of this expression. It is easy exercise to convince that if one takes $V_0 = 0$, or return to a pure Coulomb potential and substitute $P = l + 1/2$ (always valid for regular potentials), the Eq. (71) passes into (60).

From the behavior of the solution [7] at the origin it follows

$$Q_2 = \frac{512 m^5 V_0 \alpha^4 (n_r+1)(n_r+2)}{\hbar^8 (2n_r+3)^4} > 0;\ n_r = 0,1,2,\ldots P = 1 \tag{72}$$

and finally modified Heisenberg relation takes the form

$$\delta r^2 \sqrt{(\delta p^2)^2 + \frac{512 m^5 V_0 \alpha^4 (n_r+1)(n_r+2)}{\hbar^8 (2n_r+3)^4}} \geq$$

$$\geq 2\hbar \left| \frac{3}{2} + \frac{1}{(2n_r+3)} \left\{ -\frac{9}{2} + n_r \left[ \frac{33-9n_r}{3+n_r} - 3n_r - 8 \right] \right\} \right| \tag{73}$$

We see that the right side of this inequality vanishes at $n_r = 0$.

- Consider now *the singular oscillator potential*

$$V = -\frac{V_0}{r^2} + gr^2;\ (V_0 > 0, g > 0) \tag{74}$$

The standard wave function of the Schrodinger equation is

$$R = Cr^{-1/2+P} e^{-\frac{\sqrt{2mg}}{2\hbar}r^2} \left(\frac{2mg}{\hbar}\right)^{\frac{P-1/2}{4}} F\left(-n, 1+P; \frac{\sqrt{2mg}}{\hbar} r^2\right) \tag{75}$$

Energy levels are given by obvious relation

$$E = \hbar\omega(2n+1+P),\ \omega = 2\sqrt{\frac{2g}{m}},\ n = 0,1,2,\ldots \tag{76}$$

Detailed consideration of this problem is in [7], according of it

$$a_{st}^2 = \frac{4mg}{\hbar^{7/4}}(n+1) \tag{77}$$

and, therefore,

$$Q_2 = 16 m^2 \hbar^{1/4} g V_0 (n+1) > 0;\ n = 0,1,2,\ldots \tag{78}$$

If we repeat here above presented procedures for the potential (74), it follows

$$\left\langle r\frac{d}{dr} + \frac{3}{2} \right\rangle = 2n\frac{P+n}{P+n+1};\ n = 0,1,2\ldots \tag{79}$$

Here also for $n=0$ and arbitrary $l$ we derive vanishing result. For pure oscillator potential $(V_0 = 0)$ it follows

$$\left\langle r\frac{d}{dr} + \frac{3}{2} \right\rangle = 2n\frac{2l+2n+1}{2l+2n+3}; n = 0,1,2... \quad (80)$$

which vanishes again for $n=0$.

It must be noted that in [27] the following potential was considered (singular oscillator)

$$V = \frac{1}{2}kr^2 + \frac{a}{r^2} \quad (81)$$

which is going to our (67) after obvious change of notations $k/2 = g; a = -V_0; l' + 1/2 = P$ and using the solutions of this problem, the uncertainty relation reduces to direct multiplication of separate uncertainties [27]. One can check that their right-hand side is more than ours and it is no contradiction with the famous Heisenberg inequality.

Consider now one important case the Coulomb potential $V = -\alpha/r$ in the Klein-Gordon equation:

$$R'' + \frac{2}{r}R' + \left[E^2 - m^2 - \frac{l(l+1)}{r^2} + \frac{2E\alpha}{r} + \frac{\alpha^2}{r^2}\right]R = 0 \quad (82)$$

For details of solution see [28]. It is readily to calculate for $Q_2$.

$$Q_2 = 2\hbar^2\alpha^2 a_{st}^2 \lim_{r \to 0} r^{2P-2} \quad (83)$$

For non-vanishing $Q_2$ one must take $P = 1$. Therefore

$$Q_2 = 2\hbar^2 \frac{2^6 m^6 \left[\alpha^2 - (5/2 + n_r)^2\right]^3}{4!\alpha^4} \frac{(n_r+4)(n_r+3)(n_r+2)(n_r+1)}{(2n_r+5)}; n_r = 0,1,2 \quad (84)$$

Hence, the uncertainty inequality takes the form

$$\delta r^2 \sqrt{(\delta p^2)^2 + Q_2} \geq 2\hbar \left| \frac{3}{2} + \frac{1}{(2n_r+5)} \left\{ -\frac{15}{2} + n_r \left[\frac{75-15n_r}{5+n_r} - 3n_r - 12\right] \right\} \right| \quad (85)$$

where $Q_2$ is done by (83) at $P=1$. We remember that for other values of $P$, the term $Q_2$ is absent (when $P < 1$, $Q_2$ is divergent and uncertainty relation is meaningless, but for $P > 1$ - it is zero and the left-hand side becomes ordinary product of considered uncertainties). Here the r.-h.-s. vanishes when $n_r = 0$ also.

It is nothing amazing in this result, as well as the analogous results derived above about $n_r = 0$.

In [23,24] K.Urbanowski analyzed the cases of vanishing expectation values of non-vanishing commutator $[A, B]$, i.e. the cases when $\langle \phi | [A, B] | \phi \rangle = 0$. It is not necessary for $|\phi\rangle$ to be an eigenfunction of considered operators for some $|\phi\rangle \in \mathcal{H}$. Simply it may happen that for some $|\phi\rangle \in \mathcal{H}$ and some non-commuting observables $A$ and $B$ there is $\langle \phi | [A, B] | \phi \rangle = 0$. In our opinion it may happen when considered commutator is again an operator (See, comments above).

### 4. HR relation for the pair $\hat{A} = A(r,...)$ and Hamiltonian $\hat{H}_R$

Let us consider now radial Hamiltonian $\hat{B} = \hat{H}_R$ in relation (23). Following the strategy of previous discussion, we obtain

$$\int_0^\infty \hat{H}_R R^* \hat{H}_R R r^2 dr = \int_0^\infty R^* \hat{H}_R^2 R r^2 dr - i\hbar \Pi \qquad (86)$$

where [6]

$$\Pi = \frac{i}{\hbar} Y = i \frac{\hbar}{2m} \lim_{r \to 0} \left\{ r^2 \left[ \hat{A} R \frac{dR^*}{dr} - R^* \frac{d}{dr}(\hat{A} R) \right] \right\} \qquad (87)$$

For calculating this limit, the singularity of the operator $\hat{A}$ in the origin is also important. If we take

$$\hat{A} \sim \frac{1}{r^\gamma}; \quad \gamma > 0 \qquad (88)$$

Then it follows from (87) that

$$\Pi = i\hbar \frac{a_{st}^2 \gamma}{2m} \lim_{r \to 0} r^{2P - \gamma} \qquad (89)$$

and considering the regular boundary behavior $\lim_{r \to 0} R = a_{st} r^{-\frac{1}{2} + P}$ we derive

$$\lim_{r \to 0} \hat{H}_R R = A_{st} r^{-5/2 + P} \qquad (90)$$

where

$$A_{st} = \frac{a_{st}}{2m}\left[\frac{1}{4} - P^2 + l(l+1) - \frac{2mV_0}{\hbar^2}\right] \quad (91)$$

As for attractive potential $P^2 = \left(l+\frac{1}{2}\right)^2 - \frac{2mV_0}{\hbar^2}$ it follows that $A_{st} = 0$. Therefore, $Q_2 = 0$. It simply means that the Hamiltonian $H_R$ is a self-adjoint operator. Moreover, $X = 0$. Hence only $Q_1$ survives in (23).

Now we must consider various cases: $2P > \gamma, 2P = \gamma, 2P < \gamma$.

- If $2P > \gamma$ then $\Pi = \frac{i}{\hbar}Y = 0$. In this case only commutator $i\hbar\hat{C} = [\hat{A}, \hat{H}_R]$ remains in (23):

$$(\delta A)\sqrt{(\delta E)^2 + Q_1} \geq \frac{i}{2}|\langle[H,A]\rangle| \quad (92)$$

Entering here $\delta A, \delta E$ coincides with $\Delta A, \Delta E$ defined in the book of Messhiah [20].

In [6] we derived the correct relation for calculating the derivative of mean value

$$\frac{d\langle\hat{A}\rangle}{dt} = \frac{i}{\hbar}\langle[\hat{H}_R,\hat{A}]\rangle + \frac{\partial\hat{A}}{\partial t} + \Pi \quad (93)$$

(For details see [6]). In considered case $\Pi = 0$ and $\hat{A}$ does not explicitly depend on time, Eq. (93) is turned to

$$\frac{d\langle\hat{A}\rangle}{dt} = \frac{i}{\hbar}\langle[\hat{H}_R,\hat{A}]\rangle \quad (94)$$

Therefore (92) is written as

$$(\Delta A)\sqrt{(\Delta E)^2 + Q_1} \geq \frac{1}{2}\left|\hbar\frac{d\langle A\rangle}{dt}\right| \quad (95)$$

- In case of $2P = \gamma$, we have

$$\Pi = \frac{i\hbar a_{st}}{m}P = \frac{i}{\hbar}Y \quad (96)$$

and

$$(\delta A)\sqrt{(\delta E)^2 + Q_1} \geq \frac{1}{2}|\langle[H,A]\rangle + iY| \quad (97)$$

Taking into account (93) and the explicit independence on time we return to (95) again. We see that this relation does not feel the presence of $\Pi$. Same happened when $2P < \gamma$. Therefore, we can conclude that in case of central fields we need only $Q_1$. The value of $Q_1$ depends on concrete dynamics and on particular operators.

## 5. Some comments on time-energy uncertainty relation

Let us make brief comments about time-energy uncertainty relation, following to the book of A. Messhiah [20]. The author's point is based on the L.Mandelstam and I.Tamm interpretation [29] of this relation. The starting point is the HR relation (2) between any operator $\hat{A}$ and a Hamiltonian $\hat{H}$ in the form $\delta A \cdot \delta B \geq \frac{1}{2} |\langle [A, H] \rangle|$, which after taking into account the evolution equation $\langle [A, H] \rangle = i\hbar \frac{d\langle A \rangle}{dt}$ may equally well be rewritten as

$$\frac{\delta A}{|d\langle A \rangle / dt|} \cdot \delta E \geq \frac{\hbar}{2} \tag{98}$$

or else

$$\tau_A \cdot \delta E \geq \frac{\hbar}{2} \tag{99}$$

if one puts

$$\tau_A = \frac{\delta A}{|d\langle A \rangle / dt|} \tag{100}$$

$\tau_A$ appears as a time characteristic of the evolution of the statistical distribution of $A$. It is the time required for the center $\langle A \rangle$ of this distribution to be displaced by an amount equal to its width $\Delta A$, in other words, it is the time necessary for this statistical distribution to be appreciably modified. In this manner we can define a characteristic evolution time for each dynamical variable of the system [20]. It is obvious that every operator $A$ has its own $\tau_A$ time. If, in particular, the system is in a stationary state, $d\langle A \rangle / dt = 0$ no matter what $A$, and consequently $\tau_A$ is infinite; however, $\Delta E = 0$, in conformity with relation (99).

However, there is some criticism in derivation of time-energy uncertainty relations [24,30]. In accordance of these authors, there is a mathematically rigorous derivation of the position – momentum uncertainty relation and the uncertainty relation for any pair of non-commuting observables, say $A$ and $B$, but the same cannot be said about time – energy uncertainty relation, the validity of this seems to be limited. according to the common view, in quantum mechanics the time $t$ is a parameter. It cannot be described by the self-adjoint operator. Even Mandelstam-Tamm uncertainty relation (98) is also not free of controversies. Using the above described derivation of (98) authors presume that the right sides of quoted inequalities are non-zero, that is there does not exist any vector $|\phi\rangle \in \mathcal{H}$ such that $\langle [A,H] \rangle_\phi = 0$ or $d\langle A \rangle_\phi / dt = 0$. In the original paper of Mandelstam and Tamm [29] there is a reservation that for the validity of the formula of the type (98) it is necessary that $\Delta_\phi H \neq 0$ and $\Delta_\phi A \neq 0$. Hence such relations are correct only for some specific states $|\phi\rangle \in \mathcal{H}$ and observables $A$ and for others need not be correct.

However, more comprehensive analysis shows that consideration of stationary states still is possible. Indeed, for non-stationary case the uncertainty relation (95) by using (93) (for $\partial A / \partial t = 0$) can be rewritten as

$$\delta A \delta E' \geq \frac{\hbar}{2} \left| \frac{i}{\hbar} \langle [H,A] \rangle + \Pi \right| \tag{101}$$

where the redefined energy is

$$(\delta E')^2 = (\delta E)^2 + Q_1 \tag{102}$$

In non-stationary case $d\langle A \rangle_\phi / dt \neq 0$ and one can divide both sides of (101) on non-zero factor $\frac{i}{\hbar} \langle [H,A] \rangle + \Pi \neq 0$ and derive

$$\tau_A \cdot \delta E' \geq \frac{\hbar}{2} \tag{103}$$

where

$$\tau_A = \frac{\delta A}{\left| \frac{i}{\hbar} \langle [H,A] \rangle + \Pi \right|} \tag{104}$$

For stationary states, when the Hamiltonian does not depend explicitly on time and $d\langle A \rangle_\phi / dt = 0$ and owing that $\delta A \neq 0$, we can tend the denominator of (104) to zero without any problem. Therefore, $\tau_A = \infty$. Then it follows from (103) that $\delta E' = 0$, that is a correct result, because in stationary states the energy has a definite value.

On the other hand, it follows from (104) that for stationary case, since $\tau_A = \infty$, the denominator of (104) is zero. This case coincides to the hypervirial theorem discussed in [6]. So, we derive the generalized hypervirial equation in the form, proved in [6]. Therefore, if the operator $\hat{A}$ does not commute with the Hamiltonian, the generalized hypervirial theorem follows from Eq. (104) for stationary states

$$\frac{i}{\hbar}\langle[H,A]\rangle = -\Pi \qquad (105)$$

Let us check this relation in case of operator $p^2$: we have

$$[p^2, H] = 2m[H, V] = [p_r^2, V] = i\hbar(F_r p_r + p_r F_r) \qquad (106)$$

where $F_r$ is a "radial force"

$$F_r = -\frac{dV}{dr} \qquad (107)$$

and $p_r$ is a (34) Hermitian "radial momentum"

Therefore, it follows

$$\frac{d\langle p^2 \rangle}{dt} = \langle F_r p_r + p_r F_r \rangle + \Pi \qquad (108)$$

where

$$\Pi = -i\hbar a_{st}^2 \lim_{r \to 0} V' r^{2P+1} \qquad (109)$$

and so

$$\tau_{p^2} = \frac{\delta p^2}{|\langle F_r p_r + p_r F_r \rangle + \Pi|} = \frac{\delta p^2}{\left|-i\hbar\left\langle\left[2F_r\frac{\partial}{\partial r} + \frac{2}{r}F_r + F_r'\right]\right\rangle + \Pi\right|} \qquad (110)$$

It is evident that for soft-singular potentials (29) the Eq. (109) gives: $\Pi = 0$, when $P > 1$, while $\Pi$ diverges, when $P < 1$ and the time (110) is not defined. Therefore, there remains $P = 1$, when

$$\Pi = \mp 2i\hbar a_{st}^2 V_0 \qquad (111)$$

For regular potentials (28), when $P = 2l + 1$, the Eq. (109) gives nonzero result only for Coulomb potential $V = \alpha/r$ and in $l = 0$ state. But, at the same time, for Coulomb potential

$$\frac{2}{r}F_r + F_r' = 0 \qquad (112)$$

Therefore, we have

$$\tau_{p^2} = \frac{\delta p^2}{\hbar \left| 2\left\langle F_r \frac{\partial}{\partial r} \right\rangle - a_{st}^2 \alpha \right|} \quad (113)$$

For stationary state we must have

$$2\left\langle F_r \frac{\partial}{\partial r} \right\rangle - a_{st}^2 \alpha = 0 \quad (114)$$

It is impossible to satisfy to this equality without the extra term $a_{st}^2 \alpha$.
Let us check it for Coulomb potential, the radial wave function of it has a form

$$R_{10}(r) = Ce^{-\frac{r}{a_0}}; \quad C = 2\sqrt{\frac{1}{a_0^3}} \quad (115)$$

Here $a_0 = \frac{\hbar^2}{me^2}$ is the Bohr's first orbit radius for the Hydrogen atom, where $\alpha = -e^2$.

After substitution we derive the identity

$$-\frac{4e^2}{a_0^3} + \frac{4e^2}{a_0^3} = 0 \quad (116)$$

Hence, we conclude that considering the extra terms is essential to have a correct physical interpretation of the evolution time in case of stationary states.

**Conclusions**

Three-dimensional case is peculiar in many respects. It is so because the radial distance is restricted from one side, from the origin of coordinates. Therefore, it becomes necessary to take into account the boundary behavior of radial wave functions and singularity of considered operators. This fact changes some traditional expectations, which are generally valid in spaces without boundaries. Earlier we have demonstrated that some general theorems of quantum mechanics suffer distortion. The same happens in considered case of HR uncertainty relations.

Inclusion of surface terms $Q_1, Q_2, X, Y$ causes the loss of universality of the ordinary HR uncertainty relations in the sense that they become dependent on dynamical characteristics, such as potential, wave function, quantum numbers etc. These features were demonstrated especially for $r^2$ and $p^2$ pair. Moreover, for some states (quantum numbers) the corresponding uncertainty inequality reduces even to zero value on the right-hand side in spite of non-commutativity of considered operators. Like to other similar investigations this fact is not surprising [23, 24, 30]. In our case (central potential problem) the appearance of extra terms is caused by restricted domain of the radial distance in polar spherical coordinates $(0 \leq r < \infty)$.